\documentclass[showpacs,showkeys,twocolumn]{revtex4-1}
\usepackage{bm,graphicx,subfigure}
\usepackage{color}
\newcommand*\blue{\color{blue}}

\begin{document}

\title{Breathing Modes in Rotating Bose-Condensed Gas: An Exact Diagonalization Study}
\author{Mohd. Imran}\email{alimran5ab@gmail.com}
\author{M. A. H. Ahsan}
\affiliation{Department of Physics, Jamia Millia Islamia (Central University), New Delhi 110025, India.}

\begin{abstract}
We present an exact diagonalization study of the breathing mode collective excitations for a rotating Bose-Einstein condensate of $N=10$ spinless bosons interacting via repulsive finite-range Gaussian potential and harmonically confined in quasi-two-dimension. The yrast state and the low-lying excited states are variationally obtained in given subspaces of the quantized total angular momentum $L$ employing the beyond lowest Landau level approximation in slowly rotating regime with $0\le L < 2N$. For a given $L$, the low-energy eigenspectra (bands) are obtained in weakly to moderately interacting regime. Further, for a given interaction, the split in low-lying eigenenergies with increasing $L$ is the precursor to spontaneous symmetry breaking of the axisymmetry associated with the entry of the first vortex. With increase in repulsive interaction, the value of the first breathing mode increases for stable total angular momentum states $L=0~\mbox{and}~N$, but decreases for intermediate $0<L<N$ metastable states. The position of the observed first breathing modes in the eigenspectrum remains unchanged as the interaction is varied over several orders of magnitude.



\end{abstract}

\maketitle

\section{Introduction}
\label{sec:intro}
Ever since the experimental realization of Bose-Einstein condensation (BEC) in ultra-cold alkali atomic vapours in a harmonic trap \cite{aem95,dma95,bst95}, the study of collective excitations in such  systems has been an important subject of research in quantum many-body physics \cite{jem96,bdz08}.
The decisive experimental control over the density, the effective dimensionality and the atom-atom interaction strength \cite{ias98}, makes these systems an outstanding one to study subtle quantum many-body effects, in particular, collective excitations.
\\
\indent
The breathing mode (monopole oscillation or uniform radial expansion and contraction) is one of the most important collective excitations \cite{pr97,cbr02} used as a diagnostic tool for many-body effects \cite{dgp99}.
In recent years, a number of theoretical studies have demonstrated that the quantum breathing mode is ideally suited to estimate the atom-atom coupling strength, its kinetic and interaction energies and other such observables \cite{moa13}, in a trapped atomic vapour system. 
This leads to a novel kind of spectroscopy of trapped systems.
\\
\indent 
For few-body systems realized in lower dimension \cite{szl11,hhm10}, the behaviour of breathing modes can be studied to a very high degree of precision \cite{bhh02,opl10}. 
An understanding of the physics of few-body systems may then be extrapolated to larger systems to gain an insight into the beyond mean-field physics of macroscopic ensembles. 
In this note, we follow this approach and present an exact diagonalization study of the many-body effects of interaction and rotation (with quantized $L$) on the dynamics of breathing modes.

\section{Theoretical Formalism}
\label{sec:model}
We consider a system of $N$ interacting spinless bosons each of mass $M$, trapped in a harmonic potential $V({\bf r})= {\frac{1}{2}}M\left(\omega_{\perp }^{2} {r}^{2}_{\perp}+\omega_{z}^{2} {z}^{2}\right)$. The trap is subjected to  an external rotation about the $z$-axis with angular velocity ${\bm{\widetilde{\Omega}}}\equiv \widetilde{\Omega} \hat{e}_{z}$.
The trapping potential $V\left({\bf r}\right)$ is assumed to be highly anisotropic with $\lambda_{z}\equiv \omega_{z} / \omega_{\perp}\gg 1$, so that the many-body dynamics along $z$-axis is frozen. The system is thus effectively quasi-two-dimensional (quasi-2D) with $x$-$y$ rotational symmetry.
Choosing $\hbar \omega_{\perp}$ and $ a_{\perp} = \sqrt{\hbar/{M \omega_{\perp}}}$ as units of energy and length respectively, the many-body Hamiltonian in the co-rotating frame is given as
\begin{equation}
H = \sum_{j=1}^{N} \left[-\frac{1}{2} \bm{\nabla}^{2}_{j} + \frac{1}{2} {\bf r}_{j}^{2} - \Omega L\left({\bf r}_{j}\right)\right] 
+ \frac{1}{2} \sum_{i\neq j}^{N} U \left(|{\bf r}_{i}-{\bf r}_{j}|\right)
\label{mbh}
\end{equation} 
where $\Omega = \widetilde{\Omega}/{\omega_{\perp}}$ $(\leq 1)$ is the dimensionless angular velocity and $L$ (scaled by $\hbar$) is the $z$ projection of the total angular momentum operator.   
The inter-particle interaction $U$ is described by the Gaussian potential 
\begin{equation}
U \left(|{\bf r}_{i}-{\bf r}_{j}|\right) = \frac{\mbox{g}_{2}} {2\pi{\sigma_{\perp}^{2}}}
\exp{\left[ -\frac{\left(r_{\perp i}-r_{\perp j}\right)^{2}}{2\sigma_{\perp}^{2}} \right]} 
\delta \left(z_{i}-z_{j}\right)
\label{gip}
\end{equation}
with width $\sigma_{\perp}$ (scaled by $a_{\perp}$) being the effective range 
of two-body interaction.
The dimensionless parameter $\mbox{g}_{2}=4\pi {a_{s}}/{a_{\perp}}$ measures the strength of the two-body interaction with $a_{s}$ being the $s$-wave scattering length for particle-particle collision. 
We assume that the scattering length is positive $\left(a_{s}>0\right)$ so that the effective finite-range interaction is repulsive. 
The above finite-range Gaussian interaction potential is expandable within a 
finite number of single-particle basis and hence computationally feasible \cite{cfa09,dka13}. 
In the limit $ \sigma_{\perp}\rightarrow 0$, the Gaussian potential in Eq.~(\ref{gip}) reduces to the usual zero-range contact potential $\mbox{g}_{2}\delta\left({\bf r}_{i}-{\bf r}_{j}\right)$.
\\
\indent
To obtain the eigenenergies and the corresponding eigenstates, we employ exact diagonalization of the Hamiltonian matrix in different subspaces of $L$ using Davidson algorithm \cite{dav75} with inclusion of higher Landau levels in constructing the many-body basis states \cite{ahs01}.
It is to be noted that for a many-body system under consideration here, the  characteristic energy scale for the interaction is determined by the  dimensionless parameter $Na_{s}/a_{\perp}$. 
Owing to the increasing dimensionality of the Hilbert space with $N$ making the computation impractical, we vary $a_{s}$ so as to achieve the value of  $N a_{s}/a_{\perp}$ relevant to experimental situation \cite{dgp99}.

\section{Results and Discussions}
\label{sec:results}
We consider a system of $N=10$ bosonic atoms of $^{87}$Rb in a quasi-2D harmonic trap with confining frequency ${\omega}_{\perp}=2\pi \times 220$ Hz and the $z$-asymmetry parameter $\lambda_{z}\equiv {\omega_{z}}/{\omega_{\perp}}=\sqrt{8}$. 
The condensate has extension $a_{z}=\sqrt{\hbar/M\omega_{z}}$ in the $z$-direction and its dynamics along this axis is taken to be completely frozen. 
Recent advancements in atomic physics has made it possible to tune the low-energy atom-atom scattering length in ultra-cold atomic vapours using Feshbach resonance \cite{ias98}. 
Accordingly in the calculations presented here, the parameters of the two-body interaction potential in Eq.~(\ref{gip}) has been chosen as $\sigma_{\perp}=0.1$ and $a_{s}=10\, a_{0},~100\, a_{0}$ and $1000\, a_{0}$ where, $a_{0}=0.05292\,$nm is the Bohr radius. 
The corresponding dimensionless parameter $\mbox{g}_{2}={4\pi a_{s}}/{a_{\perp}}$ defined above turns out to be $0.009151,~0.09151$ and $0.9151$, respectively.
\begin{figure}[!htb]
\centering
\subfigure[$~a_{s}=10\, a_{0}$]
{\includegraphics[width=0.9\linewidth, keepaspectratio]{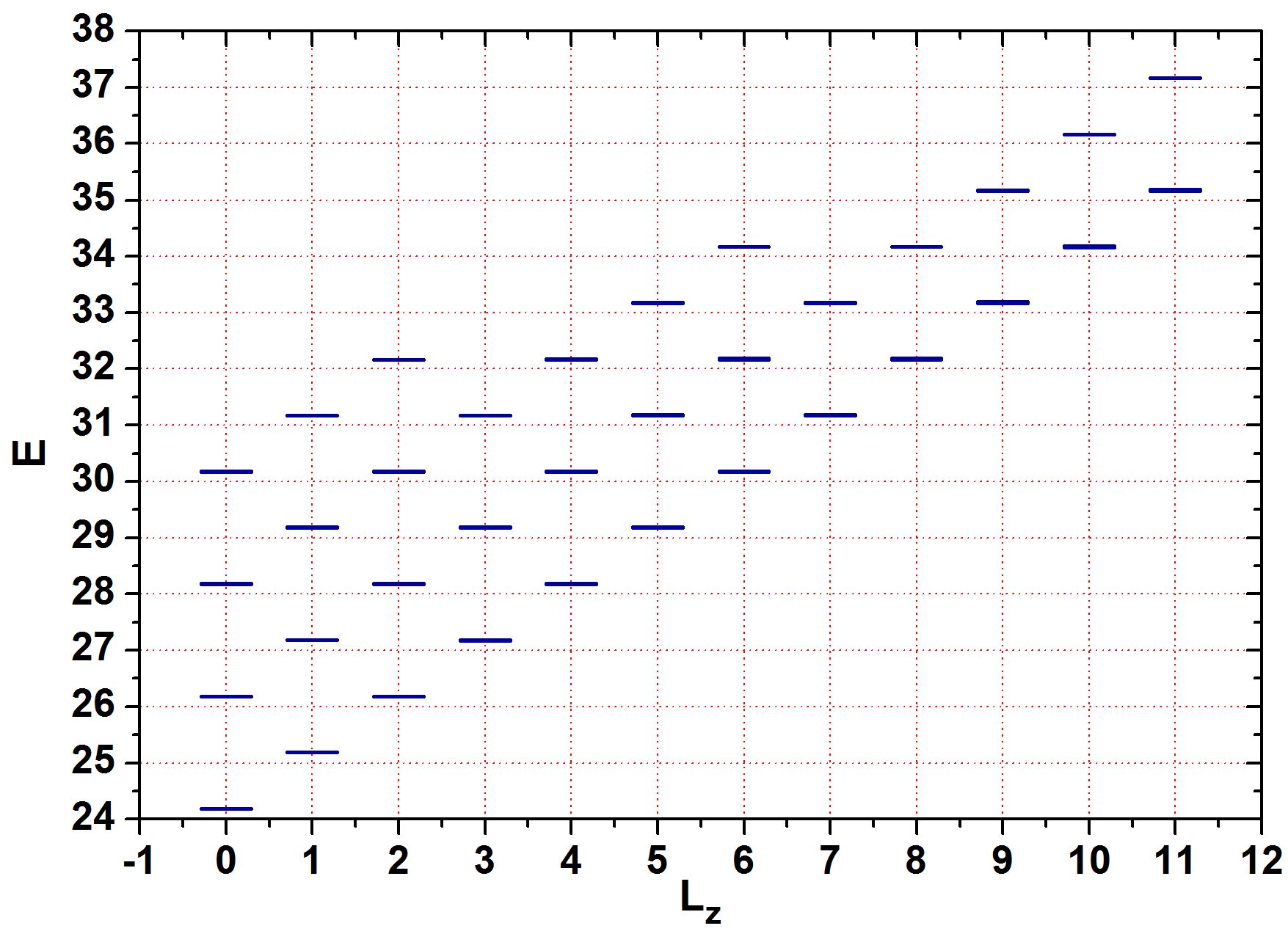}\label{sub1}}
\subfigure[$~a_{s}=100\, a_{0}$]
{\includegraphics[width=0.9\linewidth, keepaspectratio]{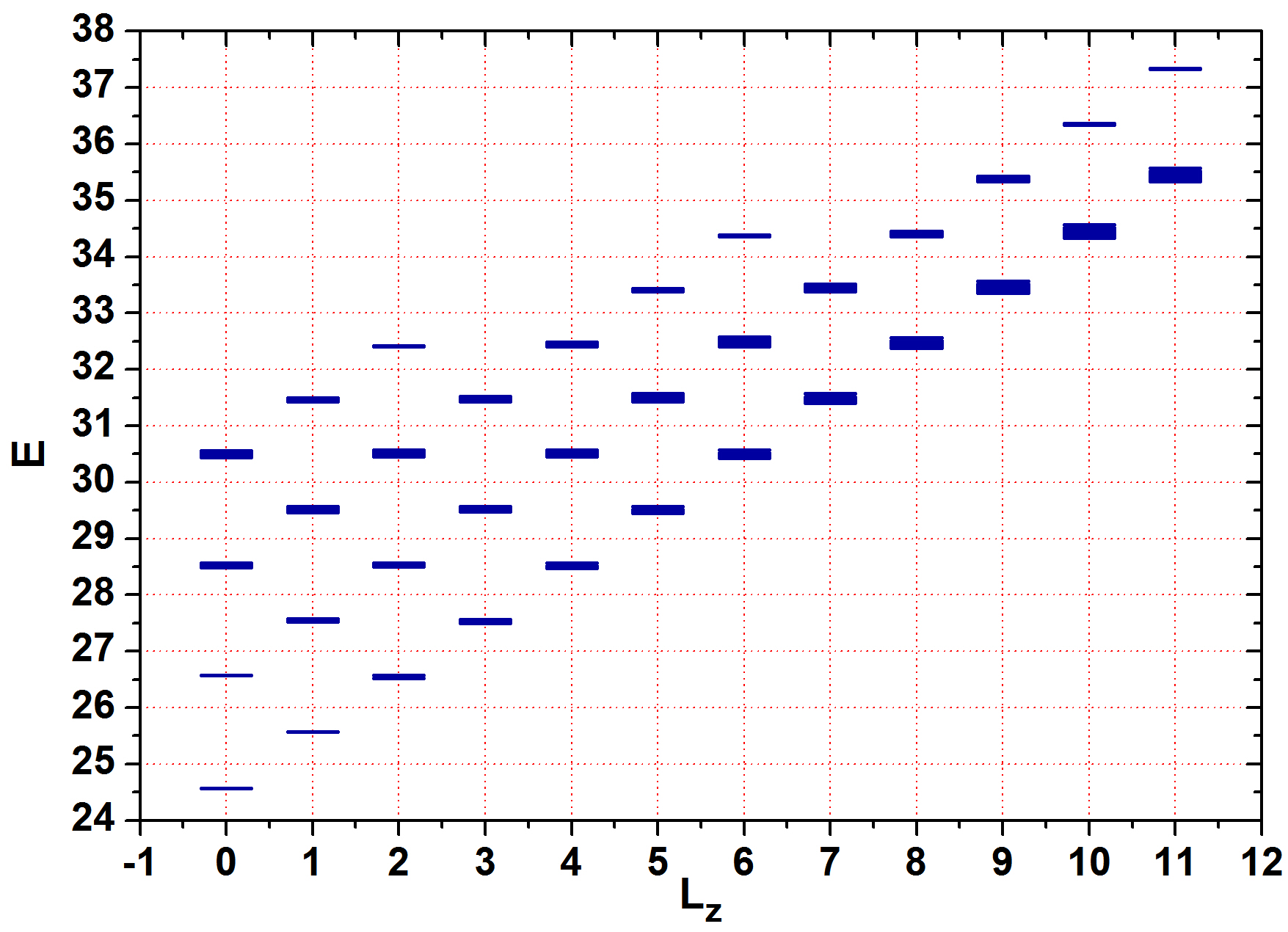}\label{sub2}}
\subfigure[$~a_{s}=1000\, a_{0}$]
{\includegraphics[width=0.9\linewidth, keepaspectratio]{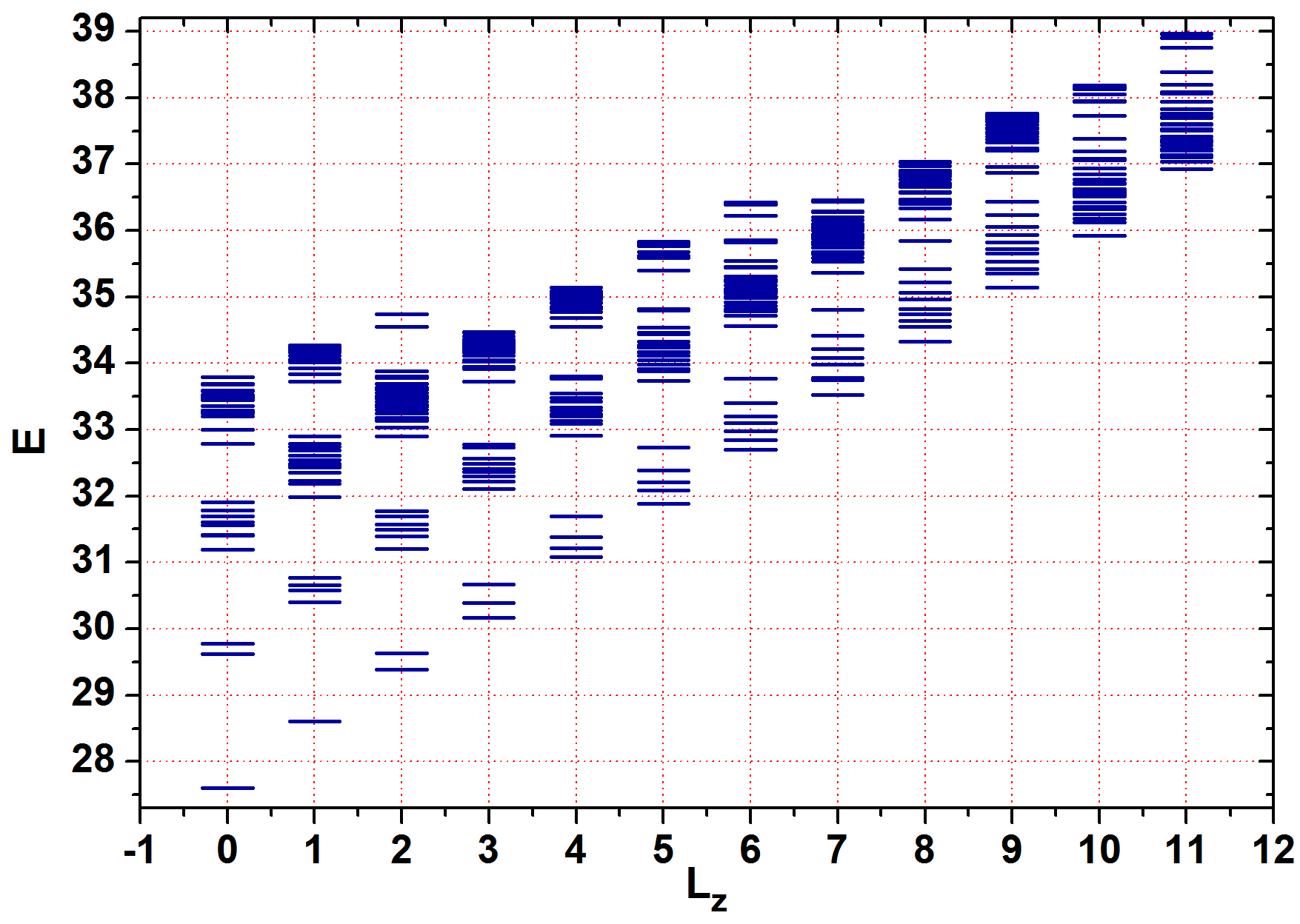}\label{sub3}}
\caption{\label{fig:esn10}For $N=10$ condensed bosons in a quasi-2D harmonic trap, the low-lying energy eigenspectrum (in units of $\hbar \omega_{\perp}$) {\it versus} the total angular momentum $L$ (scaled by $\hbar$) for three different values of repulsive interaction $\mbox{g}_{2}$ (parametrized by $a_{s}$) and $\sigma_{\perp}=0.1$ in Eq.~(\ref{gip}). In the present calculation, up to thirty low-lying energy eigenstates have been found.}
\end{figure}  
\\
\indent
Following \cite{bhh02}, we examine here the breathing modes in a rotating system of $N=10$ bosons within the quantized total angular momentum regime $0 \leq L < 2N$. 
In Fig.~\ref{fig:esn10}, we present the low-energy eigenspectra (bands) for different total angular momentum $L$ states with three different values of interaction parameter $\mbox{g}_{2}$ (parametrized by the scattering length $a_{s}$).
Eigenstates having the same total angular momenta constitute a $L$ series (or bands). 
The $i$th eigenstate of the $L$ series is denoted as $L_{i}$ and the corresponding eigenenergy as $E\left(L_{i}\right)$. 
We observe that corresponding to each $L$, the quasi-degenerate low-lying eigenenergies split to form energy bands for weakly, Figs.~\ref{sub1} and \ref{sub2}, to moderately interacting regime, Fig.~\ref{sub3}, and the respective energy gaps associated with the low-lying eigenstates are reduced. 
We also observe that for a given interaction strength, the low-lying eigenenergies further split with increase in $L$ to fill the energy gaps, as seen in Figs.~\ref{sub2} and \ref{sub3}. 
This split in low-lying eigenenergies with increasing $L$ leads to spontaneous symmetry breaking of the axisymmetry associated with the entrance of the first vortex \cite{nu03} in the angular momentum regime $0\le L\le N$.
\begin{table*}[!htb]
\caption{\label{tab:n10bm100}(Color online) The eigenenergies of the $L_{i}$ states for $N=10$ bosons with interaction parameters $\mbox{g}_{2}=0.09151$ and $\sigma_{\perp}=0.1$ in the total angular momentum regime $0 \le L \le 11$. The $L_{1}$ states are  the yrast states (or ground modes) and the $L_{i}$ states $0_{2}$, $1_{2}$, $2_{3}$, $3_{4}$, $4_{5}$, $5_{6}$, $6_{8}$, $7_{9}$, $8_{11}$, $9_{13}$, ${10}_{24}$, ${11}_{27}$ are the observed first breathing modes.}
\begin{ruledtabular}
\begin{tabular}{ccccccccccccc}
${i}$&$L=0$&$L=1$&$L=2$&$L=3$&$L=4$&$L=5$&$L=6$&$L=7$&$L=8$&$L=9$&$L=10$&$L=11$\\ \hline
1 & {\blue 24.5663} & {\blue 25.5663} & {\blue 26.5217} & {\blue 27.4976} & {\blue 28.4778}	& {\blue 29.4524} &	{\blue 30.4299} & {\blue 31.4062} & {\blue 32.3822} & {\blue 33.3583} & {\blue 34.3339} & {\blue 35.3340} \\
2 & {\blue 26.5664} & {\blue 27.5217} & 26.5665 & 27.5216 & 28.4963 & 29.4774 & 30.4416 &	31.4309 & 32.4039 & 33.3830 & 34.3578 & 35.3459 \\
3 & 26.5688 & 27.5447 & {\blue 28.4996} & 27.5670 & 28.5202 & 29.4952 & 30.4545 & 31.4361 & 32.4102 & 33.3864 & 34.3630 & 35.3563 \\
4 & {\blue 28.4833} & 27.5668 & 28.5218 & {\blue 29.4781} & 28.5675 & 29.5184 & 30.4752 & 31.4545 & 32.4308 & 33.4009 & 34.3746 & 35.3582 \\
5 & 28.5218 & 27.5687 & 28.5231 & 29.4977 & {\blue 30.4540} & 29.5679 & 30.4940 & 31.4716 & 32.4321 & 33.4109 & 34.3819 & 35.3648 \\
6 & 28.5224 & {\blue 29.4607} & 28.5392 & 29.4990 & 30.4759 & {\blue 31.4315} & 30.5164 & 31.4926 & 32.4522 & 33.4261 & 34.3830 & 35.3721 \\
7 & 28.5438 & 29.4832 & 28.5671 & 29.5051 & 30.4779 & 31.4440 & 30.5681 & 31.5143 & 32.4672 & 33.4313 & 34.3846 & 35.3769 \\
8 & 28.5448 & 29.5007 & 28.5680 & 29.5204 & 30.4807 & 31.4523 & {\blue 32.4074} & 31.5678 & 32.4911 & 33.4482 & 34.3948 & 35.3819 \\
9 & 28.5662 & 29.5018 & {\blue 30.4454} & 29.5228 & 30.4953 & 31.4557 & 32.4285 & {\blue 33.3834} & 32.5119 & 33.4627 & 34.4009 & 35.3826 \\
10 & 28.5676 & 29.5221 & 30.4598 & 29.5349 & 30.4982 & 31.4619 & 32.4324 & 33.4047 & 32.5671 & 33.4894 & 34.4056 & 35.3848 \\
11 & 28.5687 & 29.5228 & 30.4787 & 29.5661 & 30.5002 & 31.4734 & 32.4357 & 33.4060 & {\blue 34.3591} & 33.5091 & 34.4071 & 35.3921 \\
12 & {\blue 30.4398} & 29.5233 & 30.4798 & 29.5683 & 30.5184 & 31.4759 & 32.4399 & 33.4106 & 34.3808 & 33.5658 & 34.4144 & 35.3986 \\
13 & 30.4617 & 29.5365 & 30.4819 & {\blue 31.4223} & 30.5205 & 31.4779 & 32.4454 & 33.4138 & 34.3848 & {\blue 35.3347} & 34.4225 & 35.4007 \\
14 & 30.4627 & 29.5416 & 30.4990 & 31.4448 & 30.5312 & 31.4903 & 32.4529 & 33.4257 & 34.3876 & 35.3601 & 34.4279 & 35.4064 \\
15 & 30.4833 & 29.5444 & 30.4998 & 31.4557 & 30.5647 & 31.4960 & 32.4549 & 33.4306 & 34.3915 & 35.3635 & 34.4308 & 35.4127 \\
16 & 30.4836 & 29.5659 & 30.5006 & 31.4578 & 30.5686 & 31.4977 & 32.4579 & 33.4322 & 34.4033 & 35.3718 & 34.4344 & 35.4203 \\
17 & 30.4944 & 29.5676 & 30.5051 & 31.4591 & {\blue 32.4027} & 31.5161 & 32.4680 & 33.4331 & 34.4051 & 35.3814 & 34.4470 & 35.4268 \\
18 & 30.4970 & 29.5684 & 30.5153 & 31.4771 & 32.4137 & 31.5171 & 32.4724 & 33.4371 & 34.4097 & 35.3831 & 34.4530 & 35.4304 \\
19 & 30.5037 & {\blue 31.4247} & 30.5209 & 31.4779 & 32.4251 & 31.5278 & 32.4753 & 33.4407 & 34.4113 & 35.3867 & 34.4723 & 35.4317 \\
20 & 30.5040 & 31.4407 & 30.5224 & 31.4791 & 32.4330 & 31.5629 & 32.4849 & 33.4480 & 34.4127 & 35.3883 & 34.4858 & 35.4450 \\
21 & 30.5212 & 31.4456 & 30.5231 & 31.4803 & 32.4353 & 31.5687 & 32.4932 & 33.4530 & 34.4159 & 35.3914 & 34.4936 & 35.4517 \\
22 & 30.5219 & 31.4601 & 30.5238 & 31.4815 & 32.4429 & {\blue 33.3831} & 32.4956 & 33.4552 & 34.4275 & 35.3954 & 34.5161 & 35.4699 \\
23 & 30.5228 & 31.4614 & 30.5339 & 31.4916 & 32.4468 & 33.4038 & 32.5127 & 33.4624 & 34.4296 & 35.3993 & 34.5672 & 35.4854 \\
24 & 30.5229 & 31.4625 & 30.5383 & 31.4957 & 32.4534 & 33.4078 & 32.5141 & 33.4677 & 34.4305 & 35.4033 & {\blue 36.3342} & 35.4926 \\
25 & 30.5242 & 31.4676 & 30.5442 & 31.4980 & 32.4566 & 33.4090 & 32.5245 & 33.4709 & 34.4330 & 35.4069 & 36.3346 & 35.5146 \\
26 & 30.5354 & 31.4769 & 30.5642 & 31.4986 & 32.4571 & 33.4129 & 32.5608 & 33.4794 & 34.4367 & 35.4092 & 36.3466 & 35.5669 \\
27 & 30.5373 & 31.4798 & 30.5662 & 31.5006 & 32.4585 & 33.4253 & 32.5683 & 33.4899 & 34.4407 & 35.4098 & 36.3558 & {\blue 37.3287} \\
28 & 30.5416 & 31.4822 & 30.5689 & 31.5040 & 32.4619 & 33.4262 & {\blue 34.3632} & 33.4937 & 34.4490 & 35.4123 & 36.3572 & 37.3336 \\
29 & 30.5426 & 31.4834 & {\blue 32.4033} & 31.5088 & 32.4665 & 33.4297 & 34.3816 & 33.5085 & 34.4519 & 35.4203 & 36.3586 & 37.3344 \\
30 & 30.5589 & 31.4912 & 32.4232 & 31.5188 & 32.4749 & 33.4340 & 34.3843 & 33.5114 & 34.4572 & 35.4246 & 36.3599 & 37.3352
\end{tabular}
\end{ruledtabular}
\end{table*}
\\
\indent
To analyze the breathing modes, we present in Table~\ref{tab:n10bm100} the  low-lying eigenenergies $E\left(L_{i}\right)$ corresponding to angular momentum states $0 \le L \le 11$ with the interaction parameter value $\mbox{g}_{2}=0.09151$. 
The lowest energy state $L_{1}$ corresponding to an $L$ is referred to as the  yrast state (we would occasionally call it the ground mode). 
We observe that for the non-rotating $L=0$ states, $E(0_2)-E(0_1)=2.0001$ (in units of $\hbar \omega_{\perp}$). 
The states $0_1$ and $0_2$ are, thus, taken to be the ground mode (the yrast state) and the first breathing mode, respectively, for $L=0$.
It was pointed out by Pitaevskii and Rosch \cite{pr97} that a purely 2D bosonic system with zero-range interaction exhibits breathing modes arising from $SO(2,1)$ symmetry. 
The energy difference between adjacent breathing mode levels is $2\hbar \omega_{\perp}$. 
Although in the present work, the finite-range Gaussian interaction potential  has been used to replace the zero-range ($\delta$-function) interaction, the feature of $2\hbar \omega_{\perp}$ spacing is seen to persist in the energy eigenspectrum. 
Further, the $2 \hbar \omega_{\perp}$ spacing also appears for the rotating $L > 0$ states. 
For example, we find $E({1}_{2})-E({1}_{1})=1.9554$ for $L=1$, $E({5}_{6})-E({5}_{1})=1.9791$ for $L=5$ and $E({10}_{24})-E({10}_{1})=2.0003$ for $L=10$. 
Thus, the breathing modes existing in 2D systems with zero-range ($\delta$-function) interaction potential may also exist in our quasi-2D system with Gaussian interaction potential.
\\
\indent
For the non-rotating $L=0$ case (column $2$ in the Table~\ref{tab:n10bm100}), the position of the first breathing mode $0_{2}$ is always immediately above the yrast state $0_{1}$. 
However as $L$ increases, the positions (labelled by the index $i$ in column $1$) of the first breathing modes are shifted to higher eigenstates, forming a stair-like pattern of eigenenergies corresponding to the breathing modes. 
We further observe from the Table~\ref{tab:n10bm100} that the shifts in the positions of the first breathing modes are small for states with $L < N$; however, for the first vortex state with $L = N= 10$, we notice an appreciable shift in the position of the first breathing mode. 
It might be due to rearrangement of bosons on the nucleation of first central vortex. 
We have found that as the repulsive interaction is increased over three orders of magnitude, the positions of the first breathing modes in the energy eigenspectra for a given $L$, remain unchanged.
\\
\indent
We also observe the second and third breathing modes with approximately $4 \hbar \omega_{\perp}$ and $6 \hbar \omega_{\perp}$ spacing. 
For example, the states $0_{4}$, $1_{6}$, $2_{9}$, $3_{13}$, $4_{17}$, $5_{22}$ and $6_{28}$ are the second breathing modes. 
Similarly, the states $0_{12}$, $1_{19}$ and $2_{29}$ are the third breathing modes. 
Moreover as is the case for the first breathing mode, for a given $L$, the positions of these higher breathing modes are also found to be independent of the inter-particle interaction strength.
\begin{table}[!htb]
\caption{\label{tab:bmda} For $N=10$ condensed bosons, the values of the breathing modes $E_{BM}$ (in units of $\hbar \omega_{\perp}$) i.e. the difference between the eigenenergies corresponding to the breathing state ($L_{i}$) and the yrast state ($L_{1}$), for three different values of the dimensionless interaction parameter $\mbox{g}_{2}$ (parametrized by $a_{s}$) with Gaussian width $\sigma_{\perp}=0.1$ in Eq.~(\ref{gip}). With increase in repulsive interaction, the value of the first breathing mode increases for stable total angular momentum states $L=0~\mbox{and}~N$, but decreases for intermediate $0<L<N$ metastable states.}
\begin{ruledtabular}
\begin{tabular}{ccccc}
$L$&$E_{BM}$&$E_{BM}\left(10a_{0}\right)$&$E_{BM}\left(100a_{0}\right)$&$E_{BM}\left(1000a_{0}\right)$\\ \hline
 0 & $E(0_{2})-E(0_{1})$ & 1.9999 & 2.0001 & 2.0111	\\
 1 &	 $E(1_{2})-E(1_{1})$ & 1.9952 &	1.9554 & 1.7837	\\
 2 &	 $E(2_{3})-E(2_{1})$ & 1.9978 &	1.9779 & 1.8062	\\
 3 &	 $E(3_{4})-E(3_{1})$ & 1.9978 &	1.9805 & 1.9357	\\
 4 &	 $E(4_{5})-E(4_{1})$ & 1.9976 &	1.9762 & 1.8287	\\
 5 &	 $E(5_{6})-E(5_{1})$ & 1.9978 &	1.9791 & 1.8542	\\
 6 &	 $E(6_{8})-E(6_{1})$ & 1.9976 &	1.9775 & 1.8620	\\
 7 &	 $E(7_{9})-E(7_{1})$ & 1.9976 &	1.9772 	& 1.8403	\\
 8 &	 $E(8_{11})-E(8_{1})$ & 1.9976 &	1.9768 & 1.8331	\\
 9 &	 $E(9_{13})-E(9_{1})$ & 1.9976 &	1.9764 & 1.7263	\\
10 &	 $E({10}_{24})-E({10}_{1})$ & 2.0000 &	2.0003 & 2.0189	\\
11 & $E({11}_{27})-E({11}_{1})$ & 1.9994 &	1.9947 & 1.9792	
\end{tabular}
\end{ruledtabular}
\end{table}
\\
\indent
The results of the first breathing modes in subspaces of quantized $L$ are presented in Table~\ref{tab:bmda} for three different values of the interaction parameter $\mbox{g}_{2}$.  
We observe that for a given $L$, the values of the first breathing mode calculated with weak interactions corresponding to the choice $\mbox{g}_{2}=0.009151$ and $0.09151$ are very close (column $3$ and 
$4$) but significantly different from the ones calculated for moderate interaction corresponding to the choice $\mbox{g}_{2}=0.9151$ (column $5$). 
For the non-rotating $L=0$ and the single-vortex $L=N=10$ states, the values of the breathing modes {\em increase} albeit very weakly with increase in repulsive interaction. 
On the contrary for the intermediate $0<L<N$ metastable states, the values of the breathing modes are appreciably different from the non-rotating and single-vortex states and {\em decreases} significantly with increase in repulsive interaction. 
\\
\section{Conclusions}
\label{sec:conc}
In conclusion, we have examined the breathing mode collective excitations of rotating Bose-Einstein condensate with finite-range Gaussian interaction in a quasi-2D harmonic trap.
By exact diagonalization of the many-body Hamiltonian matrix in beyond lowest-Landau-level approximation, the eigenenergies of the ground and the low-lying excited states are obtained with total angular momenta $0 \leq L < 2N$, corresponding to slowly rotating regime.  
We have presented the low-energy eigenspectra (bands) of the rotating system for three representative values of different interacting regimes in subspaces of quantized $L$.
For a given $L$, the degeneracy of eigenenergies is lifted (forming energy bands) for weakly to moderately interacting regime and correspondingly the energy gaps associated with the low-lying eigenstates are reduced. 
We find that the repulsive interaction influences the values of the first breathing modes in two different ways. 
For stable $L=0,N$ states the value of the first breathing mode increases whereas for intermediate $0<L<N$ metastable states it decreases with increase in repulsive interaction.
For metastable states with angular momentum $L < N$, we observe a small shift in the position of the first breathing mode in the eigenspectrum  whereas for the vortex state $\left(L=N\right)$, the shift becomes significant.
The position of the observed first breathing mode in the eigenspectrum remains unchanged as the interaction is varied over several orders of magnitude.
 
\begin{acknowledgements}
The present work was put forth during the 4th International conference on ``Current Developments in Atomic, Molecular Optical and Nano Physics with Application" CDAMOP-2015, at Department of Physics and Astrophysics, University of Delhi, India. The authors are thankful to the organisers for providing a platform for exchange of ideas.
\end{acknowledgements}

\end{document}